\documentclass[12pt,preprint]{aastex}

\newcommand{\teff}{$T_{\rm eff}$}

\newcommand{\msun}{$M_\odot$}

\newcommand{\rsun}{$R_\odot$}
\newcommand{\logg}{$\log g$}

\shorttitle{Disks around Blue Stragglers}
\shortauthors{De Marco et al.}

\begin{document}


\title{First Evidence of Circumstellar Disks around Blue Straggler Stars\altaffilmark{1}}
          
\altaffiltext{1}{Based on observations with the NASA/ESA Hubble Space
      Telescope, obtained at the Space Telescope Science Institute, which
      is operated by the Association of Universities for Research in
      Astronomy, Inc., under NASA contract NAS5-2655.}


\author{Orsola De Marco\altaffilmark{2}, 
Thierry Lanz\altaffilmark{3,4}, 
John A. Ouellette\altaffilmark{2},
David Zurek\altaffilmark{2}, 
\& Michael M. Shara\altaffilmark{2}}

\altaffiltext{2}{Astrophysics Department, American Museum of Natural History, New York, NY 10024; orsola@amnh.org}
\altaffiltext{3}{Department of Astronomy, University of Maryland, College Park, MD 20742; tlanz@umd.edu}
\altaffiltext{4}{Visiting Scholar, Columbia Astrophysics Laboratory, Columbia University, New York, NY 10027}

\begin{abstract}
We present an analysis of
optical HST/STIS and HST/FOS 
spectroscopy of 6 blue stragglers found in the globular clusters
M~3, NGC~6752 and NGC~6397. These stars are a subsample of a set of $\sim$50
blue stragglers and stars above the main sequence 
turn-off in four globular clusters which will be presented
in an forthcoming paper. All but the 6 stars
presented here can be well fitted with non-LTE model atmospheres. 
The 6 misfits, on the other hand, possess Balmer jumps which are too large for the effective temperatures implied
by their Paschen continua. 
We find that our data for these stars are consistent with models {\it only} if we account for
{\it extra} absorption of stellar Balmer photons by an ionized circumstellar disk. Column densities of
H~{\sc i} and Ca~{\sc ii} are derived as are the the disks' thicknesses.
This is the first time that a circumstellar disk is detected around blue stragglers. The
presence of magnetically-locked disks attached to the stars has been suggested as a mechanism to lose the large
angular momentum imparted by the collision event at the birth of these stars. The disks implied by
our study might not be massive enough to constitute such an angular momentum sink, but they could be
the leftovers of once larger disks.

\end{abstract}


\section{Introduction}
\label{sec:introduction}

In the color-magnitude diagrams (CMD) of open and globular clusters (GCs), some stars appear brighter and bluer than the
main sequence turn-off, yet not as bright as stars on 
the horizontal branch (HB;
Sandage 1953; Paresce et al. 1991). Their brightnesses and colors are consistent
with main-sequence
evolutionary models for stars more massive than the cluster turnoff (e.g., Saffer et~al. 2002). 
Blue stragglers are likely to form through 
a mass-acquisition episode such as a binary merger or a stellar collision (for an overview
of stellar collisions, see Shara 2002).
In the dense environments of GCs such collision mergers should in fact be common
(Hills~\& Day 1976).

If the stragglers originate in stellar collisions, the rotation rate of the product star
should be high (Sills et~al. 2002). After an initial phase when the collision product is
bloated, a contracting phase should follow. In the absence of angular
momentum sinks, the star should spin up to faster than break-up speed and 
tear itself apart. A magnetically-locked circumstellar disk has been proposed as
an efficient angular momentum reducer (Leonard \& Livio 1995). Sills et~al. (2002) do not find evidence for the 
formation of a {\it substantial} circumstellar disk from
Smoothed Particle Hydrodynamics (SPH) simulations, including up to 
$10^6$ SPH particles, with the lightest particles having masses
of $3 \times 10^{-10}$~\msun. 

In this Letter we present the first observational evidence consistent with thin disks  
around six GC stragglers. Our conclusion follows a careful analysis that excludes other
systematic effects as potential explanations of the data. 
Whether these disks are the remnants of once more massive,
dynamically important disks, or have always been thin disks unable to serve as angular momentum
sinks, this finding is likely to provide an important window into the formation and evolution of blue stragglers.

\section{Analysis}
\label{sec:observations}

HST spectroscopy of stars in GCs M~3, NGC~6752 and
NGC~6397 was acquired 
using STIS between April 1999
and March 2001 (GO-8226),
and with FOS in October
1996 (GO-6697). 

The STIS observations adopted 
low (G430L, $\Delta \lambda$=5.5~\AA , range 3200-5600~\AA ) and intermediate (G430M, $\Delta \lambda$= 0.56~\AA , range 3800-4100~\AA ) 
resolution gratings
and the 52\arcsec$\times$0\farcs 5 slit. 
The FOS observations used
the G400H ($\Delta \lambda$= 2.82 \AA , range 3240--4822~\AA )
and G570H ($\Delta \lambda$= 4.09 \AA , range 4574--6872~\AA ) gratings with the 0\farcs 5 aperture.
Photometry of all the stars observed spectrographically was carried out on
HST/WFPC2 images. Table~1 lists 6
of the $\sim$50 stars analyzed. 

Details of the data reduction, will be presented by
De Marco et al. (2004).
Here we summarize
the extraction technique which is vital for the issues at the heart of this Letter.
For STIS observations, the size of the extraction aperture cannot be too large 
if we are to
avoid contamination of the straggler's spectrum. On the
other hand, the aperture width cannot be too small or the wings of the wavelength-dependent spatial
Point Spread Function (PSF) will not be extracted, introducing an artificial color bias.
Stars extracted with too small an aperture will be
systematically bluer in the Paschen continuum (longward of 4100~\AA ), but have
Balmer jumps and Balmer continua which are those of a cooler star. We have devoted
considerable care to
insure that the extraction signature was clearly removed from all spectra. 
The spectra are extracted with progressively wider apertures,
becoming redder as more and more
of the PSF's wings are included, up to a certain aperture width (about 8 pixels,
or 0\farcs 34). Extracting the spectra with apertures larger than this threshold
does not alter the result. Stars 
showing any significant contamination with an 8-pixel aperture have been excluded
from the sample.
For the FOS observations, the size of the entrance aperture (0\farcs 5) 
ensures that extraction problems do not affect our data.

Initial estimates of the stellar parameters were derived from a grid of LTE model
spectra calculated with Hubeny~\& Lanz (1995) spectrum synthesis code,
SYNSPEC, based on
Kurucz ATLAS9 LTE model atmospheres,
and covering the range \teff\ = 5000\,--\,25000\,K and \logg\ = 2.0\,--\,5.0.
The helium and alpha element abundances were
He/H=0.24 (by mass), and [$\alpha$/Fe]=0.3 for all clusters, following the prescription of
Bergbusch \& VandenBerg (2001). The metal abundances used were [Fe/H]=$-1.60$ for M~3 and NGC~6752 
and $-1.97$ for NGC~6397. (Harris 1996
quotes [Fe/H] = $-1.57$, $-1.61$ and $-1.95$, respectively. At these low metallicities,
differences as large as 0.1 dex do not alter the stellar parameters obtained.)
We have assumed a constant microturbulent velocity of 2~km\,s$^{-1}$. 

To compare data and model continua, we measured the fluxes at
3560, 4200 and 5450~\AA, by averaging flux values in bands 120, 80 and 120~\AA-wide,
respectively, to reduce the noise while 
avoiding absorption lines.
The fluxes were converted to magnitudes using the $U$, $B$ and $V$ 
zero-points defined by Bessell, Castelli \& Plez (1998). 
In Fig.~1 we present the comparison of models' and observations' dereddened
colors for NGC~6397. Similar diagrams will
be shown for the entire sample of $\sim$50 blue stragglers in De Marco et al. (2004). 
On these color-color plots, several stars
appear bluer (smaller values of [4200]-[5450]) than the locus occupied
by the models. For these stars, there is no model that simultaneously fits the Paschen continuum 
(and hence the [4200]-[5450] color) and the Balmer jump (represented by the [3560]-[4200] color).  
If an effective temperature is chosen to fit the Paschen continuum, the observed Balmer jump
remains too deep compared to the predicted jump by about 10\%. {\it This difference is significant,
since model fits to the other blue stragglers in the sample match the data within the noise (~$\sim$3\%).} 

In order to verify that there are no systematic errors in the theoretical colors and,
hence, to confirm that indeed there is no model that fits the data, we calculated NLTE
models with the code TLUSTY (version 198; Hubeny~\& Lanz 1995). Explicit NLTE ions
include H~{\sc i}, He~{\sc i-ii}, C~{\sc i-iii} and
Fe~{\sc i-iii}. Additional opacity sources, such as
that due to the ion H$^-$ and lines of other species are treated in LTE.
Abundances and the microturbulent velocities are assigned the same values as for
the LTE models. At these temperatures (7000$\leq$\teff$\leq$15000\,K), the NLTE
model continua and Balmer lines compare quite well with the LTE predictions.
The NLTE spectral
fits indeed confirm that there is no combination of model parameters that matches the data
for these 6 stars. In Fig.~2 we present model fits for the six stars.
The black lines represent the photospheric model spectra; the predicted jump is shallower
than the observed data (blue lines) by about $\sim$10\%. The red lines show the same models with additional absorption
by a circumstellar disk (see below). For M~3-17 the wings of the Ca~{\sc ii} lines are broad and
a $v \sin i$ = 200~km~s$^{-1}$ was adopted. The narrow core of the line is well fit
with additional absorption from a disk. For NGC~6752-11 we adopted $v \sin i$ = 50~km~s$^{-1}$,
but this is strictly an upper limit. For the other three stars, the lower resolution does not
allow to estimate stellar rotation. No interstellar contamination is expected for M~3,
because of its large heliocentric radial velocity.

Neither spectrum contamination by other stars nor reddening can explain this the Bamler observed jump discrepancy.
Blending of two stars with different
temperatures would move data points towards the {\it upper} part of the color-color diagrams (Fig.~1),
because the hotter of the two stars contributes more flux to the blue part of the spectrum
while the cooler one is a larger contributor to the red part. This results in the Paschen
continuum becoming too {\it red} (values of the [4200]-[5450] color) for any given Balmer jump;
exactly the opposite of what we observe.
We adopted $E(B-V)$ values of 0.016, 
0.056 (Schlegel, Finkbeiner \& Davis 1998) and 0.18~mag (Harris 1996), for M~3, NGC6752 and NGC~6397, respectively.
If the reddening corrections were {\em lower} than the adopted values,
the colors of our 6 stars would be
closer to those of the models and the Balmer continua would be less discrepant.
Rood et~al. (1999) determined a reddening value for M~3 of $0.02\pm 0.01$~mag, while 
Gratton et al. (2003) determined reddening values of 0.040 and 0.183~mag for NGC~6752 and NGC~6397,
respectively. From the literature, we expect reddening values' uncertainties no larger 
than 0.015~mag (NGC~6752) and 0.005~mag (M~3 and NGC~6397). 
A reddening smaller by 0.015~mag would eliminate the
Balmer continuum discrepancy for NGC6752-11, while a reddening decrement of 0.005~mag would eliminate the
discrepancy for NGC~6397-6. For the other four stars reddening decreases of 0.015~mag (M~3) and 
more than 0.020~mag (for NGC~6752 and NGC~6397) 
are required to reconcile data and models. Such relatively large reddening adjustments
are not consistent with our knowledge of reddening in the respective clusters. 
Star-to-star reddening variability over spatial scales from $< 3~\arcsec$ (NGC~6397) to $\sim$10~$\arcsec$
(M~3 and NGC~6572) are not expected to be above a few percent of the mean cluster reddening
(Thorval et al. 1997). In conclusion, while not categorically excluding reddening as a source 
of at least part of the
discrepancy, we cannot dismiss the Balmer jump discrepancy on reddening grounds.

Radiative levitation might be suspected to be
at the origin of the Balmer continuum discrepancy since it is known to
results
in large photospheric iron overabundances in horizontal-branch stars
(Moehler et al. 2003). However, our tests show that
the Balmer continuum level of the models 
is not affected significantly by an increase of the
model iron abundance by factors of 4 to 12. An
over-abundance by a factor of 4 can already be excluded for NGC6752-11,
based on the STIS intermediate-resolution spectrum. There are no indications
of metal overabundances in any of the 6 stars, suggesting that
elemental diffusion is not the cause of the Balmer jump discrepancy.

We are therefore led to conclude that the data can be 
reconciled with the models only if we allow for absorption of Balmer continuum
photons by an intervening layer of partly ionized material (Fig.~1, red lines). Stellar effective temperatures and
surface gravities were determined by fitting the Paschen continuum and the Balmer lines.
The Balmer jump is at this point poorly matched, with the models having a brighter Balmer
continuum than the data. A simple absorption model was then calculated to determine the
column density of the intervening material. An absorption model providing the necessary
continuum absorption results in Balmer line core absorptions stronger than any we observe, if the
stellar surface is completely covered by the intervening material. This suggests that the absorbing material
might be distributed in a thin circumstellar disk. In Table~2 we list 
the fraction of stellar surface eclipsed by the disk for each straggler.

The disk also absorbs \ion{Ca}{2} H\&K line photons, and we derived \ion{Ca}{2}
column densities. Except for M~3, for which can separate the interstellar
from the disk contribution because of a high radial velocity 
(--147.6~km~s$^{-1}$; Harris 1996; see Fig.~2), 
the limited spectral resolution of our data does not allow us to exclude
that the \ion{Ca}{2} absorption has an interstellar origin. 
The observed Ca~{\sc ii} K lines could be matched assuming an
interstellar column, N(Ca~{\sc ii}) $\approx 10^{12}$ cm$^{-2}$, that is consistent
with columns measured toward sources in the Galactic halo
(Smoker et al. 2003).

Finally,
the stellar radii were obtained by scaling the models' fluxes to the dereddened
PSF photometry values ($B$), while adopting $(V-M_v)$ = 
15.23, 13.26 and 12.60~mag for M~3 (Harris 1996), NGC~6752 and NGC~6397 (Gratton et al. 2003), respectively. 
From the radii, effective temperatures and gravities, we determined 
the bolometric luminosities and the masses (5 out of 6 masses are convincingly larger than 
the clusters' turn-off masses; this will be fully discussed in the context of the complete sample
by De Marco et al. [2004]).
Our results and their uncertainties are summarized in Table~2.

\section{Discussion}
\label{sec:discussion}

From the Boltzmann equation, the disk electron temperatures
must typically be larger than 9000~K in order to populate the first excited level of hydrogen.
The derived column densities depend moderatly on the disk temperatures, which were assumed to be
the same as the stellar effective temperatures.
From the Saha equation, the hydrogen ionization fraction is 10$^{-3}$ - 10$^{-4}$, implying a
collisionally-ionized disk. The ratio of the \ion{Ca}{2} to \ion{H}{1} column densities is typically
1$\times$10$^{-4}$, which is
consistent with the stellar Ca/H abundance ratio when we account for the hydrogen ionization
fraction and assume no dust depletion of calcium. 

With a {\it total} H column density of 10$^{25}$\,cm$^{-2}$ (Table~1), assuming a disk radius of 0.1~AU
(larger disk radii should be rare in the crowded environments of GC centers), and
a thickness of 0.03~AU derived from 
a covering factor of 20\% and a stellar radius of 2.0~\rsun , from simple geomentry
we obtain a total H density of 1$\times$10$^{-11}$~g~cm$^{-3}$ and a total mass of 2$\times$10$^{-8}$~\msun . 
An increase of the disk radius increases the disk mass proportionally.

We suggest that {\it most} stragglers might have circumstellar disks. The disk pushes data points in the color-color diagram
down (Fig.~1). Only those data points that shift outside the locus covered by the models will be recognized as stars with
disks. For those stars whose disks do not change their colors enough to displace them outside the locus covered by the models, 
an alternative model (cooler and with lower gravity) will be found that fits the data.
A method of detecting these disks might be by UV spectroscopy of low ionization metal line.

The disks implied by our analysis are irrevocably non-massive. Although typical SPH simulations
could not be used to predict disks as light as these, the one-million-particle
simulation of Sills et al. has particle masses as small as $\sim$10$^{-10}$~\msun\ and might have forecast the presence of
disks as light as $\sim$10$^{-8}$~\msun. On the other hand, magneto-hydrodynamical transport of angular momentum, not included 
in the SPH code of Sills et al., might
be important. More work is needed both observationally and
theoretically to understand the presence and role of disks in blue straggler formation.

\acknowledgments

OD acknowledges support from Janet Jeppson Asimov.

%\appendix

%\section{Appendicial material}

\clearpage
%\begingroup
                                                                                                                                          
\begin{deluxetable}{lll}
\tabletypesize{\small}
\tablecaption{Observations}
\tablewidth{0pt}
\tablehead{
\colhead{Star} & \colhead{RA \& Dec (J2000)} & \colhead{Datasets} }
\startdata
M3-17    & 13 42 10.6  +28 22 58.6 & O5GX130(10,20)\tablenotemark{a} \\
N6397-4  & 17 40 42.3  -53 40 30.5 & Y3FJ010(B,C)T\\
N6397-5  & 17 40 42.2  -53 40 32.1 & Y3FJ010(L,M,N)T\\
N6397-6  & 17 40 42.4  -53 40 32.7 & Y3FJ010(P,Q)T\\
N6397-7  & 17 40 41.3  -53 40 25.5 & Y3FJ010(4,5)T \\
N6752-11 & 19 10 52.0  -59 59 07.0 & O5GX180(10,20)\tablenotemark{a}\\
\enddata
\tablenotetext{a}{The star spectra are at y-pixel 772 and 103, respectively}
%\tablenotetext{b}{The star spectrum is at y-pixel position 103}
\end{deluxetable}

%\endgroup

%\clearpage
\begingroup
                                                                                                                                                             
\catcode`?=\active
\def?{\phantom{1}}
                                                                                                                                          
\begin{deluxetable}{lllllll}
\tabletypesize{\small}
\tablecaption{Stellar and disk parameters}
\tablewidth{0pt}
\tablehead{
\colhead{Parameter} & \colhead{M3-17} & \colhead{NGC6752-11} & \colhead{NGC6397-4}  & \colhead{NGC6397-5} &
\colhead {NGC6397-6} & \colhead {NGC6397-7}}
\startdata
$E(B-V)$ [mag]          &0.016  &0.056 &0.18    &0.18   &0.18 &0.18\\
\teff\ [K]              &10\,000&9000  &13\,000 &10\,200&10\,000 &10\,500\\
$\log$ (g [cm~s$^{-2}$])&3.8    &3.5   &4.2     &4.2    &4.0  &3.7\\
$R$ [\rsun]             &2.4    &3.4   &1.0     &1.9    &1.9  &2.7 \\
$M$ [\msun]             &1.35   &1.30  &0.62    &2.03   &1.29 &1.30\\
$v \sin i$ [km~s$^{-1}$]&200$\pm$50 & $<$50 & \nodata & \nodata & \nodata & \nodata\\
$U$ [mag]               &16.09  &13.92 &15.21   &14.53  &14.61&13.64\\
$B$ [mag]               &\nodata&13.87 &15.36   &14.66  &14.71&13.79\\
$V$ [mag]               &16.06  &13.94 &15.40   &14.59  &14.63&13.82\\
\smallskip
$\log$ (N(H~{\sc i}))   &20.8   &21.7  &19.8    &20.9   &21.0 &20.7\\
$\log$ (N(Ca~{\sc ii})) &16.0   &17.3  &17.0    &17.0   &17.0 &17.0\\
covering fact.           &0.25   &0.20  &0.20    &0.15   &0.25 &0.15\\
\enddata
\tablecomments{Uncertainties: $\Delta$\teff=250~K; $\Delta\log g$=0.1;
The covering factor is uncertain by 0.05; The column densities $N$ are uncertain by 0.2~dex; 
Distances are assumed to be accurate to 10\%, as is the compound uncertainty 
of scaling the stellar model flux to the data. The resulting uncertainty on the masses is thus typically 50\%.}
\end{deluxetable}
                                                                                                                                         
\endgroup

\clearpage

\begin{figure}
\vspace{6cm}
\includegraphics{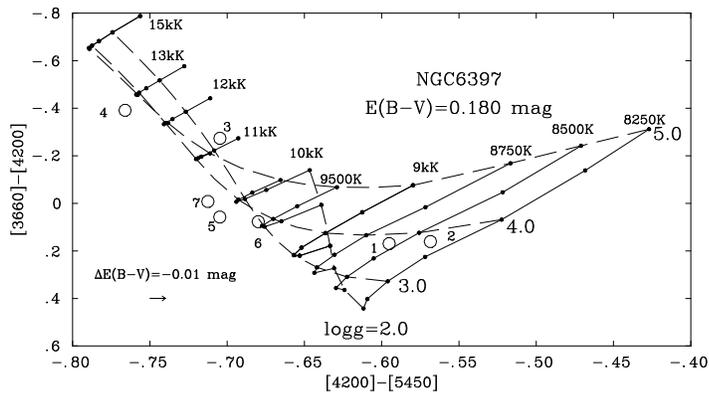}
\caption{Color-color diagrams using monochromatic magnitudes comparing spectrophotometric data
for GC NGC~6397 (open symbols) and Kurucz, LTE
models (small filled symbols). Solid lines join models of equal effective temperature, while dashed lines
join models of equal gravity (labelled). 
The adopted reddening is
labelled as is the shift in color due to a decrease in reddening of 0.01~mag. For stars 1, 2 and 3
an excellent fit to the data can be found (with no disks). These fits will be presented by
De Marco et al. (2004).
\label{fig:col_col}}
\end{figure}

\clearpage

\begin{figure}
\vspace{21cm}
\includegraphics{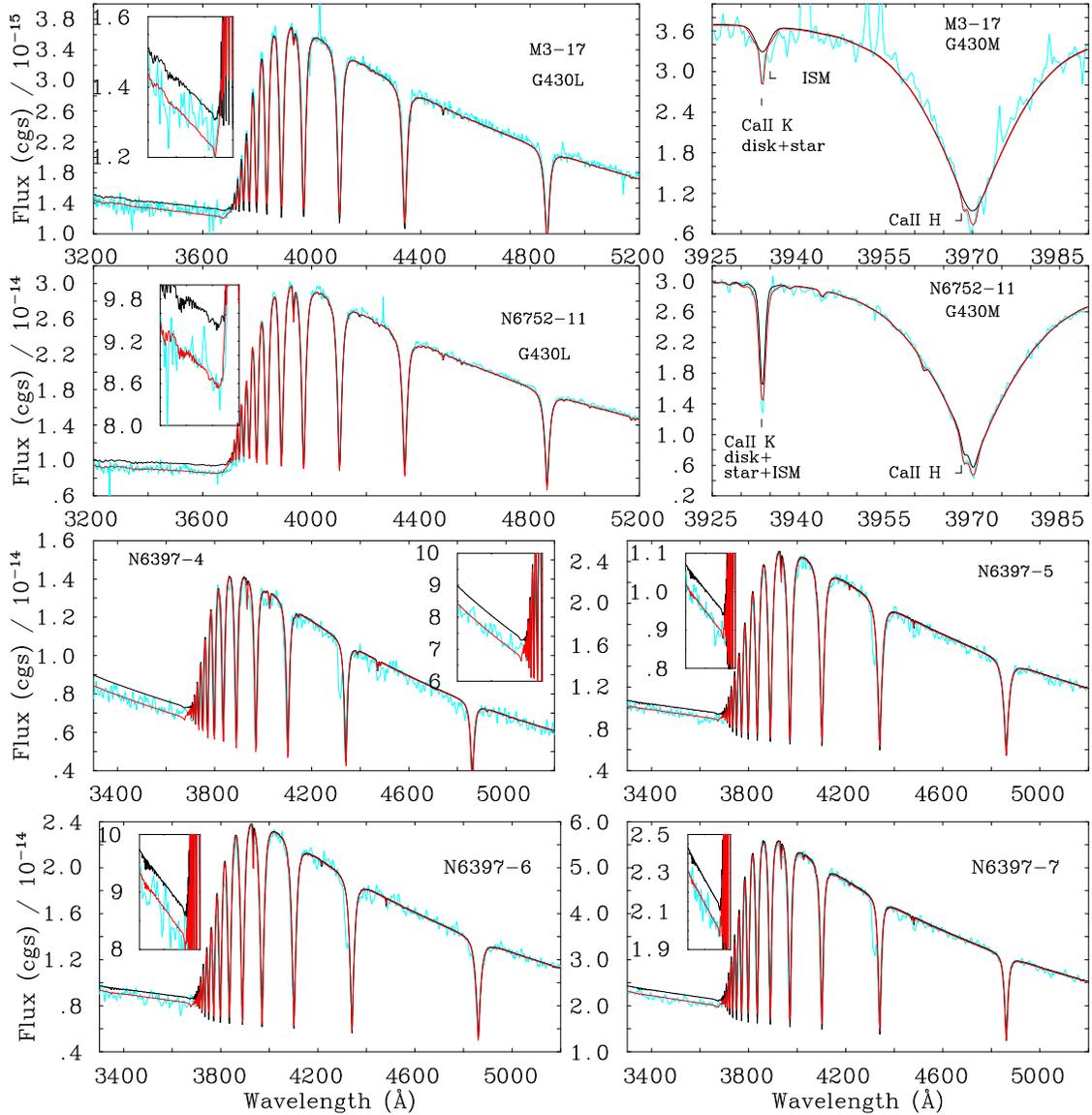}
\caption{NLTE model fits (black lines) to 6 blue stragglers (blue lines). There is no model that 
can fit the Paschen and Balmer continua simultaneously. Absorption from an ionized disk
orbiting each blue straggler
(red lines) reconciles the models with the data both in the continuum and lines.
\label{fig:disks}}
\end{figure}
\end{document}